\documentclass[10pt]{iopart}

\usepackage{graphicx}
 \expandafter\let\csname equation*\endcsname\relax
  \expandafter\let\csname endequation*\endcsname\relax
\usepackage{amsmath}
\usepackage{setspace}
\usepackage[numbers]{natbib}
\usepackage{url}

\renewcommand{\thesection}{\Roman{section}}
\renewcommand{\thesubsection}{\thesection.\Alph{subsection}}
\renewcommand{\thesubsubsection}{\thesubsection.\arabic{subsubsection}}

\begin{document}

\title{Intrinsic pinning by naturally occurring correlated defects  in FeSe$_\text{1-x}$Te$_\text{x}$ superconductors}
\author{M.\,  L.\,  Amig\'o$^1$, \, \,  M.\,  V.\,  Ale Crivillero$^1$, \, \,  D.\,  G.\,  Franco$^1$\footnote{Present address Max-Planck-Institute for Chemical Physics of Solids, D-01187 Dresden, Germany}, \, \,   A.\,  Bad\'ia--Maj\'os$^2$,    \,J.  Guimpel$^1$,  \, J. Campo$^2$, \,   F.  Damay$^3$, \, \,    F. \, Porcher$^3$, \, \,    A. \, Cond\'o$^1$  \, \,and  \, \, G.  \,Nieva$^1$}
\address{$^1$Centro At\'omico Bariloche (CNEA) and Instituto  Balseiro (U. N. Cuyo), 8400 Bariloche, R\'{i}o Negro, Argentina.}
\address{$^2$Dep. F\'isica de la Materia Condensada--Inst. Ciencia de Materiales de Arag\'on, Universidad de Zaragoza--CSIC, E-50009 Zaragoza, Spain}
\address{$^3$Laboratoire L\'eon Brillouin, CEA-CNRS UMR 12, 91191 Gif sur Yvette Cedex, France}
\ead{gnieva@cab.cnea.gov.ar}
%\date{\today}
\vspace{10pt}
\begin{indented}
\item[]February 2017
\end{indented}

\begin{abstract}
We study the angular dependence of the dissipation in the superconducting state  of  FeSe and Fe(Se$_\text{1-x}$Te$_\text{x}$)  through electrical transport measurements, using  crystalline intergrown materials.
 We reveal the key role of the inclusions of the non superconducting magnetic phase Fe$_\text{1-y}$(Se$_\text{1-x}$Te$_\text{x}$), growing into the Fe(Se$_\text{1-x}$Te$_\text{x}$) pure $\beta$-phase, in the development of  a correlated defect structure. 
The matching of both atomic structures defines the  growth habit of the crystalline material as well as the correlated planar defects orientation. 

\end{abstract}

%\vspace{2pc}

\maketitle
\ioptwocol
\section{Introduction}\label{introduction}

Within the Fe-based superconductors, the PbO type Fe chalcogenides, i.e.: the ``11" compounds, have the simplest structure  but a very complex electronic and magnetic behaviour\cite{kasahara2016giant}.
In particular, the normal and superconducting states of  Fe(Se$_\text{1-x}$Te$_\text{x}$) are strongly influenced by  hydrostatic and chemical pressure\cite{sun2016dome, mizuguchi2010review}. Furthermore, due to the very rich thermodynamic phase diagram\cite{schuster1979transition}, slight differences in material synthesis yield a variation in composition and the appearance of  defects that  significantly change the electronic and magnetic properties.
The emergence of the hexagonal and magnetic (NiAs-type structure) Fe$_7$Se$_8$   intergrowth  in the (PbO-type tetragonal)  $\beta$-FeSe matrix,  has been the source of  confusion in earlier reports of superconducting and normal properties of these materials. A low temperature synthesis is needed to grow clean single crystals of the tetragonal  $\beta$ phase of these compounds\cite{chareev2013single}.
Despite the above, the intergrown samples are very useful to understand the effect on the normal state electronic properties of the magnetic ordering within the non magnetic matrix, and also of the  microscopic stress induced changes in the normal state properties and in the superconducting parameters in the mixed state. 
On the other hand, the microscopic geometry of the intergrown phases structural coupling determines the kind of defects (geometry and distribution) that influence the vortex dynamics in the dissipationless state and the properties of the dissipation below the superconducting critical field and above the zero critical current state.  
In a certain range of defect density, the potential application of these materials in wires or tapes, has to be considered. These intergrown crystals are possible emergent materials suitable for application given their high superconducting critical field and the simple way to generate correlated defects  in them.

This article is focused on the correlated defects in superconducting  FeSe and in the Fe(Se$_\text{1-x}$Te$_\text{x}$) family. We study the microscopic nature of these defects characterizing local and bulk transport properties. Correlation is established by combination of a number microscopic and macroscopic experimental techniques in a wide range of applied temperatures and magnetic fields. This has allowed us to separate the intrinsic and defect induced contributions to superconductivity.

\section{Crystal growth and characterization}\label{crystal growth}

Fe$_\text{1-{y}}$(Se$_\text{1-x}$Te$_{x}$) crystals  of nominal composition x = 0, 0.25, 0.5, 0.75 and 1  with an intergrowth of two phases were obtained by a multi-step method. The starting materials were $>$4N powders  of Fe, Se and Te, ball milled in  Ar atmosphere.
The obtained Fe(Se$_\text{1-x}$Te$_\text{x}$) powders  were sealed in an evacuated quartz tube and fired at 680\,$^\circ$C  for 48 hours, in order to obtain a polycrystalline material. Low heating rates were used while the temperature was close to the melting point of Se and Te and  the cooling rates were free in all cases.
 Afterwards, Fe$_\text{1-{y}}$(Se$_\text{1-x}$Te$_\text{x}$)  crystals were grown in evacuated double quartz ampoules using a 1/4 KCl:3/4 NaCl flux, and a  molar proportion Fe(Se$_\text{1-x}$Te$_\text{x}$):flux of 1:50. The upper temperature reached depended on the composition of the crystals, increasing with the Te content. For x = 0, 0.25, 0.5, 0.75 and 1
the final temperatures were 850, 868, 885, 903 and 920\,$^\circ$C respectively. The cooling rates were 0.05\,$^\circ$C/min from the final temperature to 600\,$^\circ$C.  At 400\,$^\circ$C the quartz tubes were quenched in water. Finally, the crystals were released dissolving the flux with water.

Platelet-like shaped crystals were obtained. As explained below, the microstructure was characterized by room temperature X-ray (XR) diffraction and transmission electron microscopy (TEM).  Additional structural studies of the powders,  so as to assess a number of physical parameters (characteristic structural and magnetic transition temperatures), were performed by neutron thermo-diffraction (see Appendix for details).  The crystal composition was  measured by Energy Dispersive Spectroscopy (EDS).  Angle resolved transport measurements were performed with a conventional four-wire method in a superconducting magnet up to 16\,T. For such investigation, the crystals were mounted onto a rotatable sample holder with an angular resolution of 0.05\,$^\circ$. Magnetization and superconducting volume fraction of the samples were determined using a QD-SQUID magnetometer with a the zero field cooling protocol. 

\begin{figure}[b]%h de here, b de bottom y t de top
\begin{center}
\includegraphics[angle=90,origin=c]{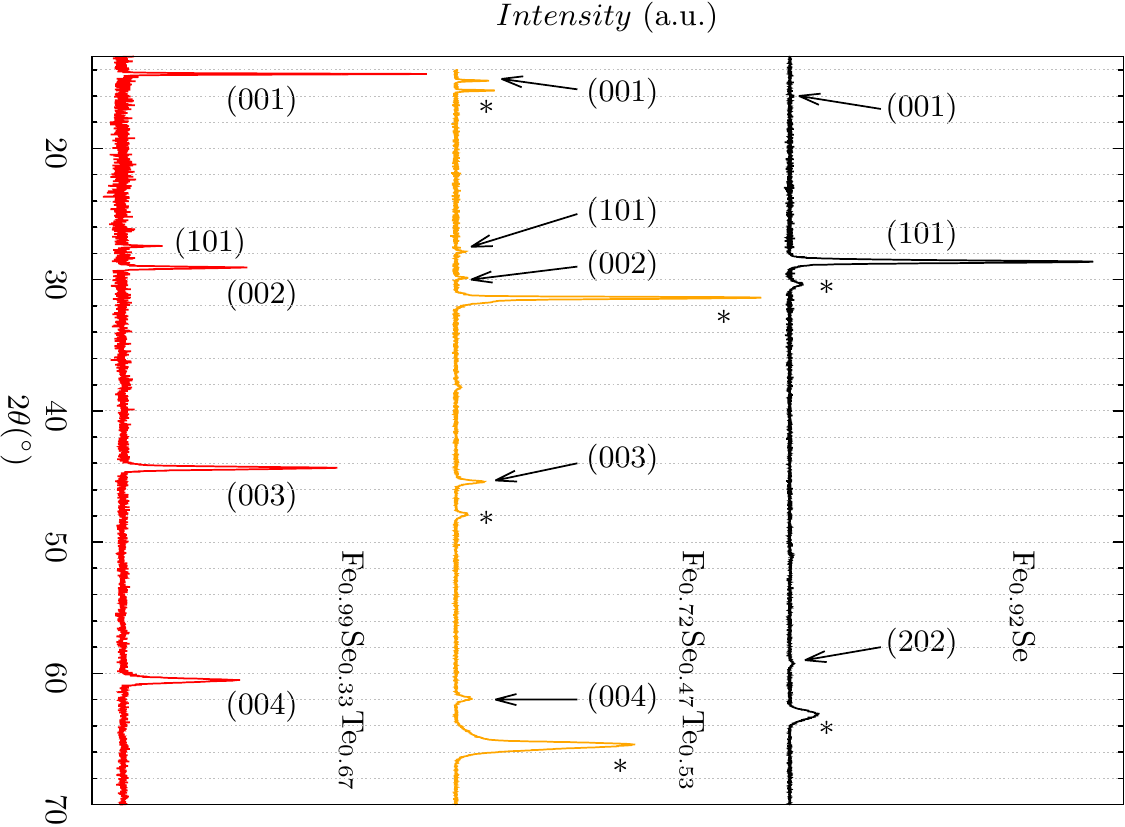}
\end{center}
\caption{XR-diffraction patterns for Fe$_{0.92}$Se, Fe$_{0.68}$(Se$_{0.47}$Te$_ {0.53}$) and Fe$_{0.99}$(Se$_{0.33}$Te$_ {0.67}$). Numbers in brackets indicate the Miller indices for the tetragonal phase. $\ast$ indicates peaks corresponding to the $(00l)$  family of the  hexagonal phase.}
\label{Rx}
\end{figure}
The resulting crystalline samples show Fe deficiency as  indicated by EDS. Samples without Te have an average composition of Fe$_{0.89}$Se, while crystals containing Te have a lower average Fe concentration: $(1-{y}) \simeq 0.78$. This is an indication that the crystals have a mixture of structural phases because pure $\beta$-phase with a tetragonal PbO type structure would be near y=0. It is therefore expected to find inclusions of  NiAs type hexagonal phase. 
Indeed, the XR diffraction data of our crystals show the presence of both tetragonal and hexagonal phases. Other authors have found nanoscaled hexagonal phase in a tetragonal matrix and that the shape and size of this intergrowth depends on the crystal growth speed \cite{wittlin}. Figure \ref{Rx} shows XR diffraction patterns for Fe$_\text{1-{y}}$(Se$_\text{1-x}$Te$_\text{x}$) crystals, with the incidence plane of the XR perpendicular to the platelet's surface. We show results for samples with the lower and the higher Fe content. 

For Fe$_{1-{y}}$Se, we identified the tetragonal PbO  type structure with the space group $P4/nmm$ of the $\beta$ phase and the hexagonal NiAs type structure of the Fe$_7$Se$_8$ (Fe$_{0.875}$Se) compounds. In the tetragonal phase, we found the  reflections from  $(00l)$ and $(l0l)$ planes families, while in the hexagonal phase, only the $(00l)$ reflections were present. For the samples containing Te, we observed that those with  lower Fe content have larger hexagonal phase peaks, and that the $(l0l)$ reflections are comparatively less intense.

\begin{figure}[t]%h de here, b de bottom y t de top
\begin{center}
\includegraphics[width=8cm]{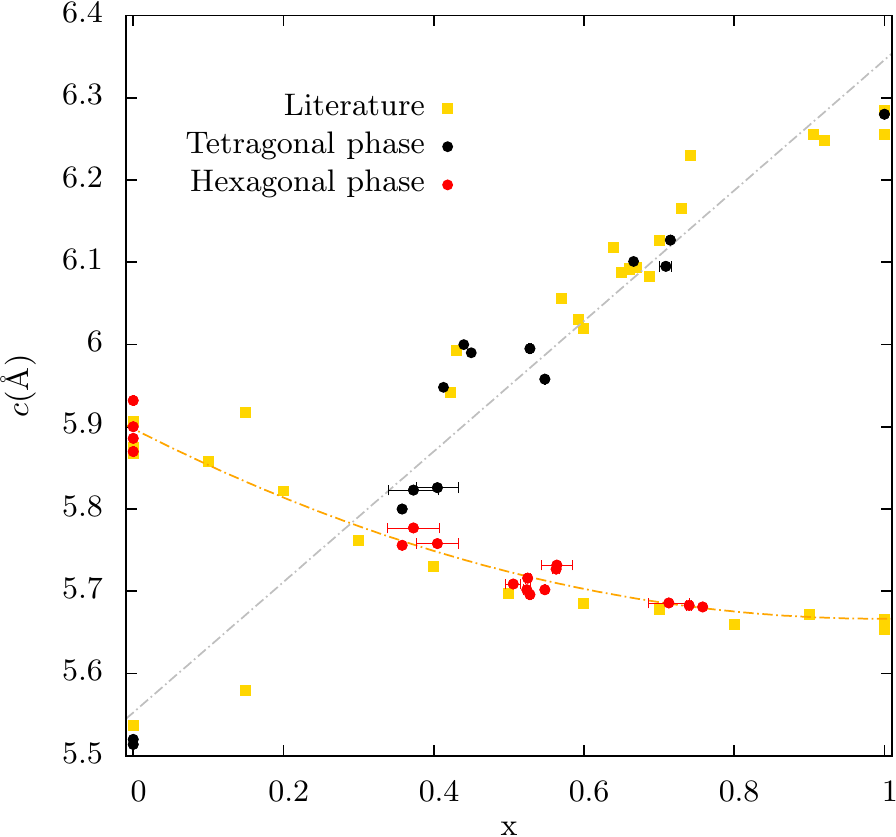}
\end{center}
\caption{$c$ lattice parameter of the tetragonal and hexagonal phases as a function of the content of Te. Square symbols show data  from the literature \cite{Gawryluk, Sales, Argy, Gresty, fang-eje-c, FeTe, Fe7Se8-ejec, zhang, terzieff1981magnetism, yeh2009superconducting}.}\label{ejec}
\end{figure}
Figure \ref{ejec} shows two different behaviors of the $c$ axis parameter as a function of  Te concentration, corresponding to the tetragonal and hexagonal phases. 
 The $c$ axis of the tetragonal phase increases with x, in good agreement with the measurements of other authors \cite{Gawryluk, Sales, Argy, fang-eje-c, Gresty, yeh2009superconducting}. On the other hand, the $c$ axis of the hexagonal phase, decreases with the content of Te. 
We also show in Figure \ref{ejec} that our hexagonal phase $c$ axis data coincide with those taken by Terzief\cite{terzieff1981magnetism} who studied the structural properties of the solid solution Fe$_\text{1-{y}}$(Se$_\text{1-x}$Te$_\text{x}$) between Fe$_7$Se$_8$ and  Fe$_2$Te$_3$. It's worth mentioning that, in
 spite of the Te ions being bigger than the Se ones, $c$ axis decreases with increasing Te content. According to our observations, this is due to the increase in the amount of Fe vacancies and to the fact that the structural parameters of the hexagonal phase change smoothly  and continuously from  Fe$_7$Se$_8$ to Fe$_2$Te$_3$. 

\begin{figure}[b]%h de here, b de bottom y t de top
\begin{center}
\includegraphics[width=8cm]{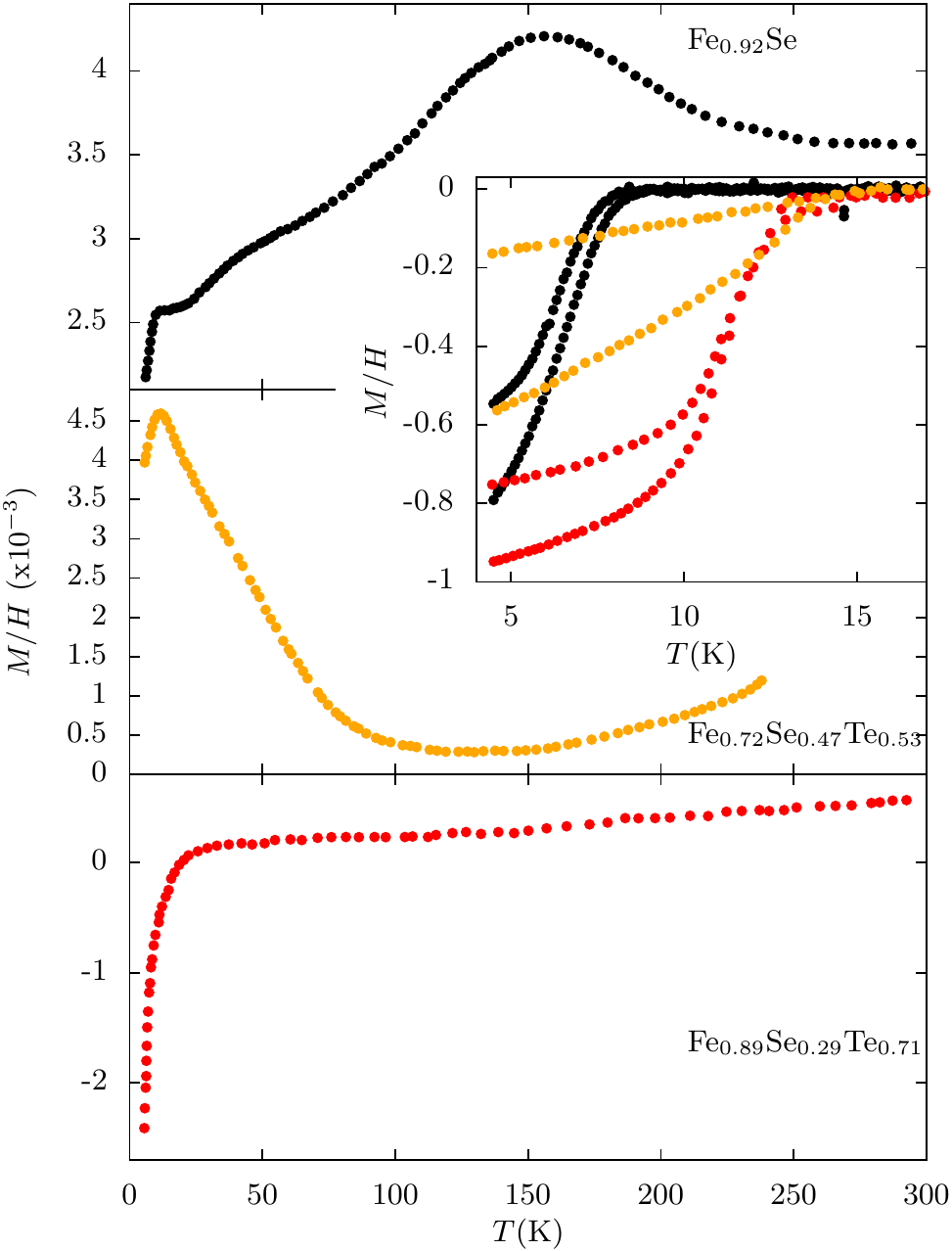}
\end{center}
\caption{Temperature dependence of the magnetization for Fe$_{0.92}$Se, Fe$_{0.72}$(Se$_{0.47}$Te$_{0.53}$) and Fe$_{0.89}$(Se$_{0.29}$Te$_{0.71}$)  crystals with a 0.1\,T applied magnetic field perpendicular to the crystal plane. $Inset:$ Superconducting volume fraction, measured with $H=0.2$\,mT.}\label{MvsT}
\end{figure}

In the magnetization measurements it is expected to find characteristics of both phases present in our crystals. Figure \ref{MvsT} shows the temperature dependence of the magnetization with an applied magnetic field of 0.1\,T perpendicular to the crystal plane for different Te concentrations. At low temperatures, there is a drop in the magnetization due to the superconducting transition. Due to the presence of the hexagonal phase, the behavior in the normal state depends on the Te concentration. 
As described in the literature\cite{terzieff1981magnetism}, concomitant Fe vacancies are located in alternate planes and as long as they are ordered, the material is magnetic. Depending on the value of y, hexagonal Fe$_\text{1-{y}}$Se, is ferrimagnetic (FM), antiferromagnetic (AF) or has a transition from FM to AF\cite{terzieff1978antiferromagnetic}. 
 Fe$_7$Se$_8$ has two types of vacancy superstructures called 3c and 4c. They both have a FM behavior at high temperatures and at  $\sim$130\,K or  $\sim$220\,K they have a change in the easy axis of magnetization causing a drop in the magnetization (AF)\cite{japoneses}.

In figure \ref{MvsT},  Fe$_{0.92}$Se has a broad hump at $\sim$150\,K that we associated with the presence of hexagonal phase, Fe$_7$Se$_8$ with vacancies superstructure,  in agreement  with our magnetic neutron diffraction study presented in the Appendix.

With increasing Te content, the hump rapidly disappears as expected if the hexagonal phase content decreases.
 In the case of Fe$_{0.72}$(Se$_{0.47}$Te$_{0.53}$), the system tends to order at temperatures below 100\,K but does not saturate. In the samples with more Te, the system is paramagnetic. In the inset of figure \ref{MvsT} the susceptibility as a function of temperature below the superconducting onset is shown  for a magnetic field of H=0.2\,mT perpendicular to the crystal plane. We obtain a superconducting volume fraction of 80\%, 56\% and 94\% for Fe$_{0.92}$Se, Fe$_{0.72}$(Se$_{0.47}$Te$_{0.53}$) and Fe$_{0.89}$(Se$_{0.29}$Te$_{0.71}$), respectively.

\section{Results and discussion}\label{results}

Figure \ref{rvst}a) shows the typical behavior of the in plane resistivity as a function of temperature for crystals of Fe$_\text{1-{y}}$(Se$_\text{1-x}$Te$_\text{x}$). As analyzed below, due to the coexistence of the tetragonal and hexagonal phases, and depending on the  value of the resistivity, $\rho$,  of each phase  and on the domain size of each phase,   $\rho$  might present characteristics of both phases.

\subsection{Normal state properties}\label{normal}

The main features of the normal state resistivity in samples of the $\beta$-phase have been described in the literature\cite{fang, sudesh, liu, McQueenRx}. A metallic like behavior was reported for the Se rich end.
The Te doping in samples with only $\beta$-phase presents a metallic behavior below a certain $T^\ast$.  Excess Fe in these samples  results in a weak localization regime.  In the Te rich end,  a marked magnetic and structural anomaly at $\simeq$70\,K that separates two different resistivity behaviors was reported.
\begin{figure}[t!]%h de here, b de bottom y t de top
\begin{center}
\includegraphics[width=8.5cm]{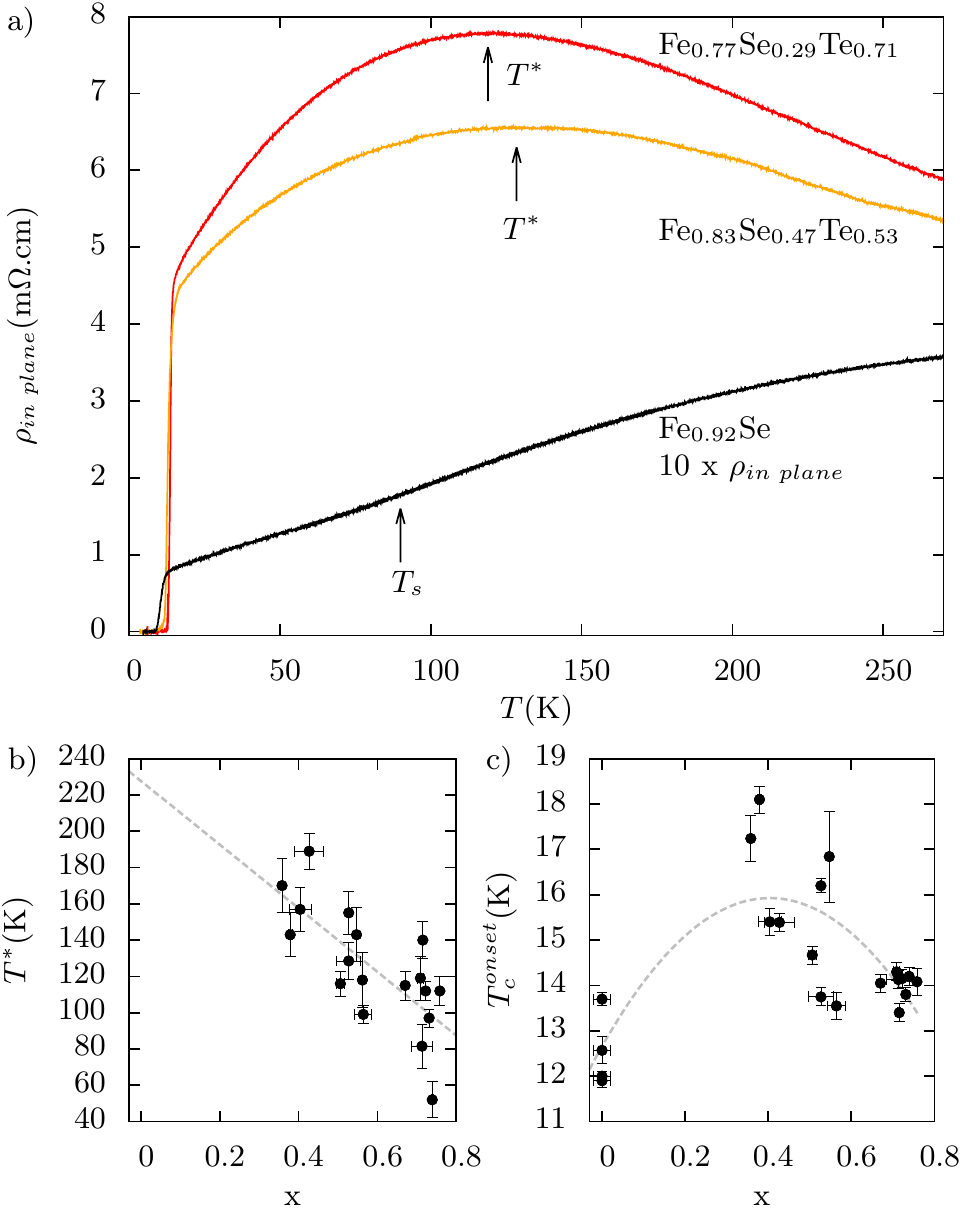}
\end{center}
\caption{a) Temperature dependence of the in-plane resistivity for Fe$_{0.92}$Se, Fe$_{0.83}$(Se$_{0.47}$Te$_{0.53}$) and Fe$_{0.77}$(Se$_{0.29}$Te$_{0.71}$) crystals. $T_s$ indicates the temperature of the structural transition and $T^\ast$, the maximum in the resistivity. b) $T^\ast$ and c)  $T^{onset}_c$ as a function of the Te content. The dotted lines are guides to the eye.}\label{rvst}
\end{figure}
In Fe deficient compounds, the resistivity is metallic-like in Fe$_7$Se$_8$ with an important drop at $T\simeq$60\,K which is related\cite{jap-resist} to a magnetic reorientation at $T\sim$130\,K while Fe$_2$Te$_3$ presents a semiconducting behavior\cite{Aramu1964}.

The in plane resistivity of our Fe$_\text{1-y}$Se  crystals exhibits a metallic-like behavior down from room temperature with a change in the slope at $T_s \simeq 90$\,K. This change is associated with a structural transition of the $\beta$-phase from tetragonal to orthorhombic\cite{mcqueen-phase-trans}. Our neutron diffraction experiments on powders (see Appendix) clearly support this fact.
In Te doped samples, the resistivity increases with decreasing temperature until a certain $T^\ast$ at which it becomes metallic-like. This is a common feature of all of our samples, and increasing the Te content, $T^\ast$ decreases as shown in figure \ref{rvst}b). On the other hand, $T^\ast$ has no dependence on the Fe content within the range studied in this work. Moreover, we have found no evidence of weak localization as reported in Ref. \cite{liu}. 
The behavior above $T^\ast$ is unlikely to be related to the semiconducting hexagonal phase Fe$_{2}$Te$_3$ because it is also seen in pure $\beta$-phase crystals, both in the literature\cite{sudesh} and in our crystals. It does not have a clear and unique interpretation but it  may be related with the maximum in resistivity of $\beta$-FeSe at 350\,K\cite{0953-2048-28-10-105009}. Furthermore, the Te dependence of $T^\ast$ has the same behavior as the pressure dependence of the resistivity maximum in Fe(Se$_{0.5}$Te$_{0.5}$) \cite{doi:10.1143/JPSJ.78.063705} where $T^\ast$ increases with pressure. In our case, Te replacement by Se is equivalent to a positive pressure that is reflected in an increase of $T^\ast$ with Se content.

\subsection{Superconducting behavior}\label{superconducting}

As we have stated in the previous section, the lower the amount of Fe the greater the proportion of hexagonal phase, but this varies with the Se content. This fact is also made evident in the superconducting transition. Our Fe$_\text{1-{y}}$Se crystals present a transition around $T^{onset}_{c} \simeq $12\,K, higher than the transition temperature for pure $\beta$-phase samples  ($T_c$= 8.5\,K)\cite{lour}. 
As customary, the transition temperature is defined by the intersection of the linear extrapolation from the normal state and the transition zone.

The possibility of stress  at the interfaces between the hexagonal and tetragonal phases and in consequence increment of the transition temperature is discussed  in the Appendix. In brief, a systematic widening of the neutron diffraction peaks for the samples with mixture of phases, corresponds to the existence of such stresses, that have an influence on the pressure dependent $T_c$. On the other hand, the Te dependence of the transition temperature presents a maximum at x$\simeq$0.4 as is expected from other results in the literature\cite{katayama2010investigation}, see figure \ref{rvst}c). 

We will now discuss the superconducting state of Fe deficient Fe$_\text{1-{y}}$(Se$_\text{1-x}$Te$_\text{x}$) crystals analyzing the effect of the microscopic mixture of phases in the vortex state dissipation in transport experiments. Given the tetragonal layered crystalline structure these compounds are expected to present anisotropic electronic properties that will manifest also in the mixed state.

\subsubsection{Intrinsic anisotropy ($H_{c2 \parallel} / H_{c2 \perp}$)}\label{intrinsic}
\begin{figure}[b]%h de here, b de bottom y t de top
\begin{center}
\includegraphics[width=8.5cm]{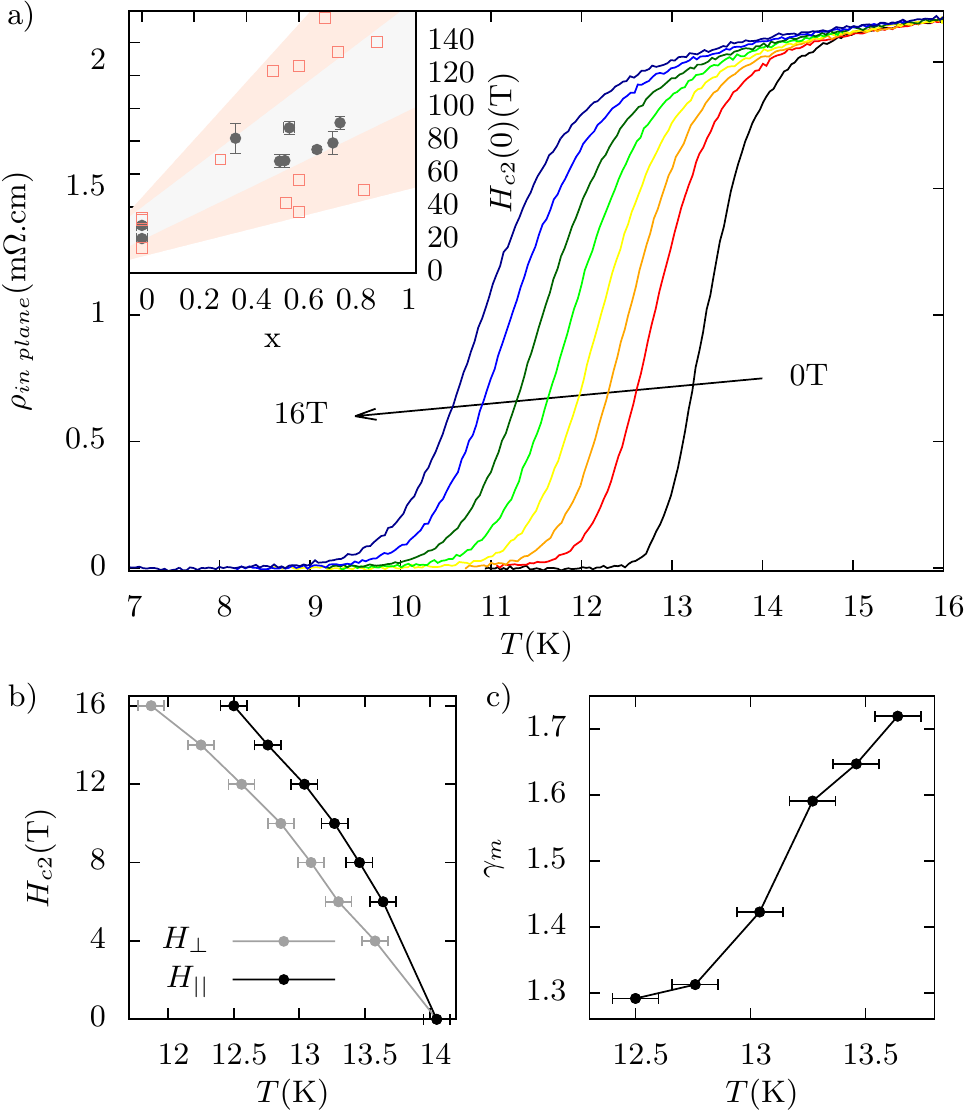}
\end{center}
\caption{a) Temperature dependence of the in plane resistivity for a Fe$_{0.99}$(Se$_{0.33}$Te$_{0.67}$) single crystal  with $H_\perp$ (0, 4, 6, 8, 10, 12, 14 and 16\,T). The arrow show the direction of field  increase. $Inset:$ Te composition dependence of  $H_{c2\perp}$(0) for our data (grey circles) and values extracted from the literature\cite{lei, Ge, kida2010weak, kida2009upper, khim, Yadav, lei2012multiband, Hc2}. b) Phase diagram for $H_{c2 \parallel}$ and $H_{c2 \perp}$. c) Temperature dependence of the anisotropy $\gamma_{m}$. } \label{rvstSC}
\end{figure}

\hfill\\
Figure \ref{rvstSC}a) shows the in plane transition for a Fe$_{0.99}$(Se$_{0.33}$Te$_{0.67}$) single crystal at various magnetic fields perpendicular to the crystal surface, $ H_\perp$. Similar behavior is observed for other compositions. The transition is broadened by about 15\% from 0 to 16\,T. This broadening is consistent with a thermal activated behaviour of the vortex lattice, see section \ref{broadening}.

The superconducting phase diagrams for fields parallel, $ H_{||}$, and perpendicular, $H_{\perp}$, to the platelet's plane are shown in figure \ref{rvstSC}b). The definition of $ H_ {c2}$ is given by the temperature of the onset of the transition for each applied field  defined as the intersection between the linear extrapolations from normal and transition zones. The difference  between both orientations  makes evident that the material is slightly anisotropic. The ratio $\gamma_{m}$ = $H_ {c2 \parallel } / H_ {c2 \perp} $ indicates that the anisotropy increases with temperature in agreement with other works \cite{lei, fang}, see figure \ref{rvstSC}c). For Fe$_\text{1-{y}}$Se, the anisotropy between in plane and out of plane $H_ {c2}$ measurements also increases with temperature, with $\gamma_m$ being approximately 1.2. However, as discussed in previous sections,  the superconducting $\beta$-phase has two main orientations perpendicular to the planar surface of the samples (i.e. $00l$ and $l0l$), therefore making the measured value of  $\gamma_m$  lower than the value for the clean  $\beta$-phase, found to be approximately 3\cite{lour2}.

Using the conventional one-band WHH theory \cite{lei}, 
\begin{equation}\label{WHH}
\ H_{c2}(0)=-0.693 \frac{{d}H_{c2}}{{d}T} \Big|_{T_c} T_c,
\end{equation}
we obtained the value of $H_ {c2\perp}$(0) as a function of  composition. The inset of figure \ref{rvstSC} shows the comparison of our data with those extracted from the literature\cite{lei, Ge, kida2010weak, kida2009upper, khim, Yadav, lei2012multiband, Hc2}. As reported by other authors for samples with Fe excess\cite{Hc2}, in our Fe deficient samples it can be observed  that also  $H_{c2}$(0) increases with the content of Te.

\subsubsection{Broadening of $\rho (T)$--Activated behavior}\label{broadening}
\begin{figure}[h]%h de here, b de bottom y t de top
\begin{center}
\includegraphics[width=8.5cm]{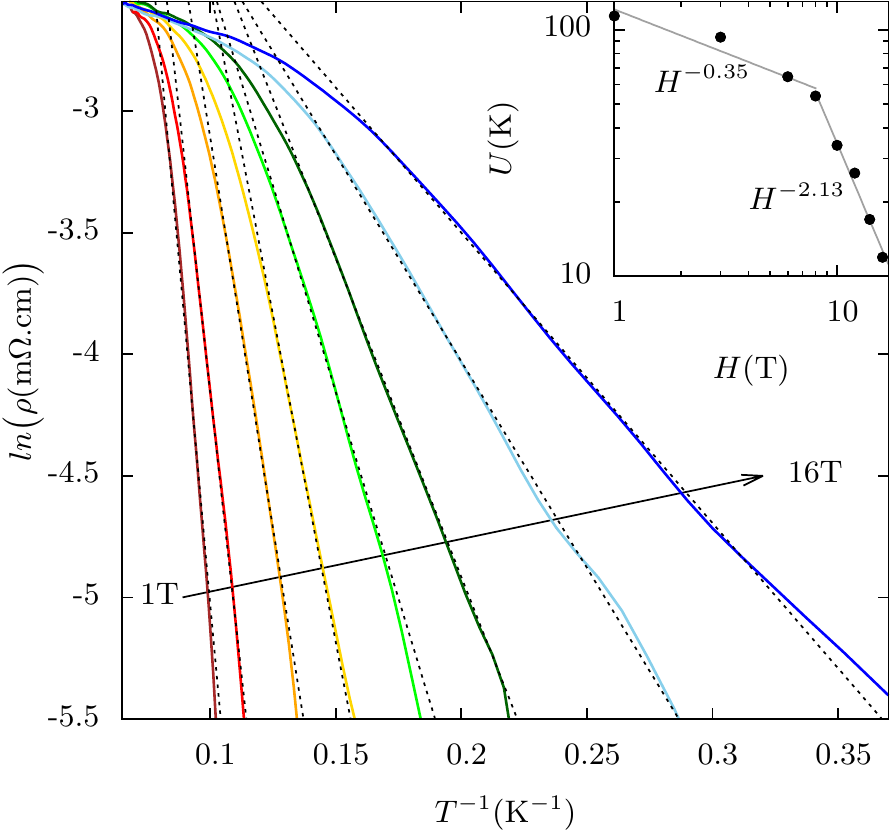}
\end{center}
\caption{Logarithm of the in plane-resistivity as a function of $T^{-1}$ for Fe$_{0.91}$Se. The
dotted lines are the fit of equation \ref{rhovsU} for each magnetic field ($H$=1, 3, 6, 8, 10, 12, 14, 16\,T). $Inset:$  Activation energy for the vortex movement as a function of the applied magnetic field. The solid line is the fit of $U_{0} H^{-q}$.
}\label{rvsU}
\end{figure}

\hfill\\
Figure \ref{rvsU} shows the logarithm of the resistivity vs  $T^{-1}$, for several applied magnetic fields perpendicular to the crystal surface. The linear behavior over a wide temperature range suggests  that  a thermally activated flux motion model\cite{palstra} suitably encodes the broadening of the transition in Fe$_{0.91}$Se.
Thus, in order to describe the dissipation below $H_{c2}$ in the region where the macroscopic critical current goes to zero, $J_c\to 0$, we used such activated flux motion model\cite{palstra}.

In the linear response regime, the resistivity is well approximated by,
\begin{equation}\label{rhovsU}
\ \rho= \rho_0 exp\Big(\frac{-U(H,T)}{T}\Big),
\end{equation}
with the activation energy
\begin{equation}
\ U(H, T)=  U_0 H^{-q} \Big(1-\frac{T}{T_c}\Big).
\end{equation}
Here, $T$ is the temperature, $H$ is the applied magnetic field,  
$U_0$, %is the activation energy for the vortex movement
 $\rho_0$ and $q$ are constants and $T_c$ is the transition temperature at $H=0$\,T. 
Figure \ref{rvsU} also shows the fits to the data.
The activation energy extracted from the slope of these fits is shown on the inset. The two values of the $q$ exponent for the lowest and highest fields are found in several works\cite{Yadav, Ge, palstra88} but they do not have a clear interpretation.

%\break

\subsubsection{Angular dependence $\rho (\theta)$ in Fe$_\text{1-y}$(Se$_\text{1-x}$Te$_\text{x}$).}\label{angular}%\hfill\\
The interpretation of the results in the previous paragraphs (Sec.\ref{intrinsic} and Sec.\ref{broadening}) requires to consider the following framework:  the dissipative regime observed in our samples may be related both to the electronic anisotropy of the material and to the nature of the defects in the structure. Close to the normal state, the resistivity is concomitant to intrinsic superconducting properties (higher critical field, mass anisotropy%, \dots
) Below a certain $\rho/\rho_n$ value, that depends on the nature of the defects\cite{figueras2003influence}, the dissipation is affected by an effective viscosity that hinders the movement of vortices, thereby affecting the activated behavior. Hence, by means of the activated movement picture one may obtain an effective activation energy that gives information  on the nature of the defects as pinning centers.

\begin{figure}[b]%h de here, b de bottom y t de top
\begin{center}
\includegraphics[width=8cm]{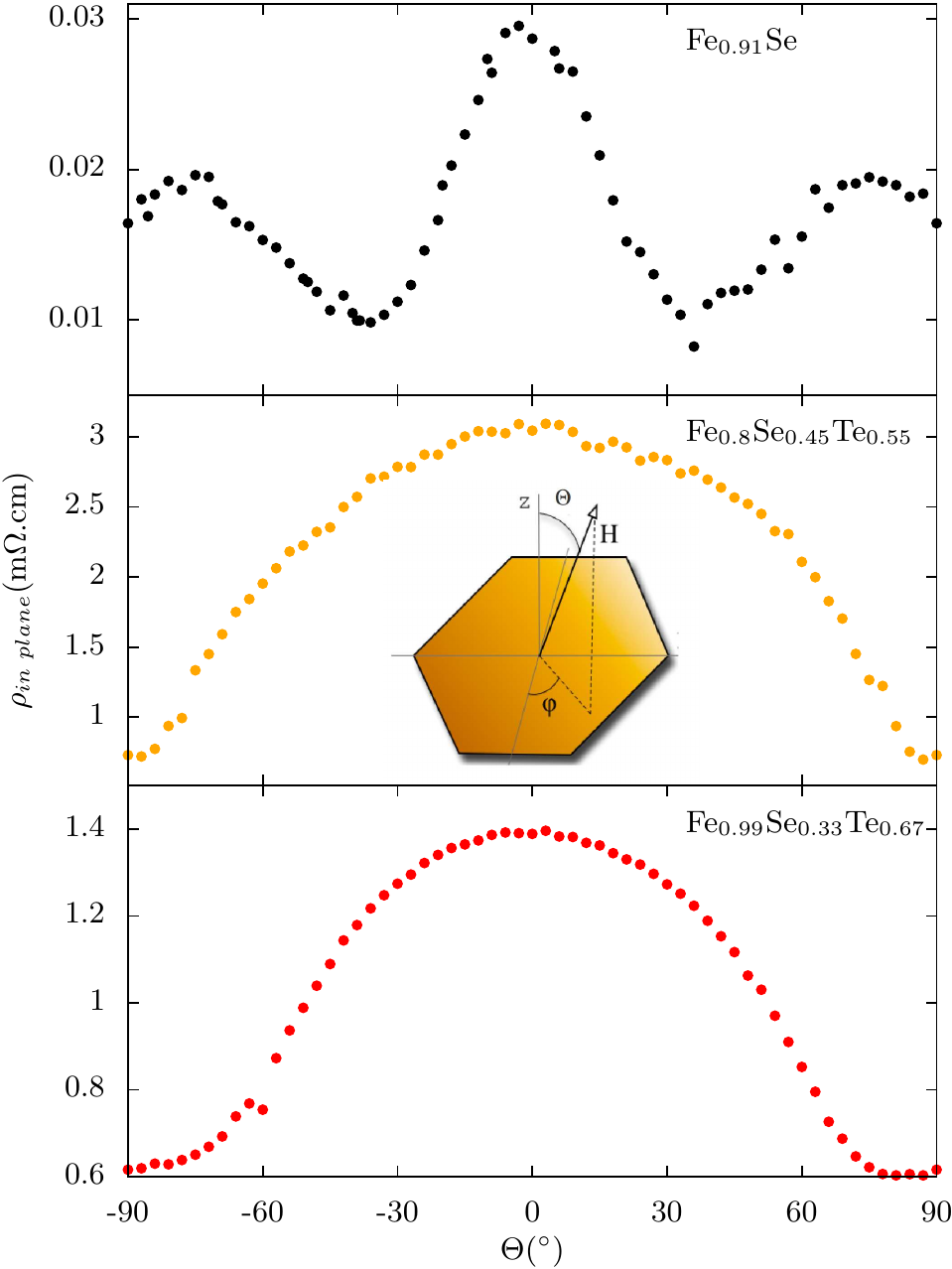}
\end{center}
\caption{Angular dependence of the in-plane resistivity for Fe$_{0.91}$Se, Fe$_{0.8}$(Se$_{0.45}$Te$_{0.55}$) and Fe$_{0.99}$(Se$_{0.33}$Te$_{0.67}$)  single crystals  with $H$=12\,T and $T$=6.70(3)\,K,
$T$=13.00(7)\,K and $T$=11.80(7)\,K, respectively. The $ inset$ is a sketch of the sample with the $\Theta$ angle definition. The in plane applied current is always perpendicular to the field.}\label{dep-ang}
\end{figure}

Then, in order to study the influence of possible correlated defects, we measured the angular dependence of the dissipation in transport experiments below the onset of the superconducting state.
We used a constant  Lorentz force configuration, maintaining the measuring current along the rotation axis of the crystal. The perpendicular to the crystal's surface is 0$^\circ$ in our notation.

Figure \ref{dep-ang} shows the angular dependence of the in plane resistivity, $\rho_{in\ plane}$, for Fe$_{0.91}$Se, Fe$_{0.8}$(Se$_{0.45}$Te$_{0.55}$) and Fe$_{0.99}$(Se$_{0.33}$Te$_{0.67}$) for $H$=12\,T and $T$=6.70(2)\,K, $T$=13.00(7)\,K and $T$=11.80(7)\,K, respectively. For each sample, the temperature is chosen so that $\rho(T) / \rho(T_c) \sim 0.5$. 

According to the model\cite{blatter} of an anisotropic material with point defects, the resistivity, at a fixed $H$ and $T$, should have a maximum at 0\,$^\circ $, a minimum at 90\,$^\circ$ and behave monotonically between these two extremes.
To be specific, the angular dependence\cite{blatter}  should be given by, $\rho(\Theta,H,T)=\rho(H\varepsilon_\Theta,T)$, where $\varepsilon^2_\Theta=cos^2 \Theta + \gamma ^{-2} sin^2 \Theta$, and $\gamma$ is the mass anisotropy of the material.
In our case, crystals with Te display this behavior but with small discrepancies in the slope of the curves approaching the orientation of 90\,$^\circ$. Such discrepancies do not relate to the Te or Fe concentration. On the other hand, Fe$_{1-y}$Se presents two deep minima at $\Theta \sim \pm 34\, ^\circ$. These extra minima were found in all the samples in a wide range of fields and temperatures below $T_c^{onset}$. Furthermore, for the larger crystals, these $34\,^\circ$ minima appear independently of the direction of the current in the crystal plane and on the contact arrangement (not shown here). In the literature,  similar measurements of angular dependence for FeSe\cite{patel} have been reported.

\begin{figure}[t]%h de here, b de bottom y t de top
\begin{center}
\includegraphics[width=8cm]{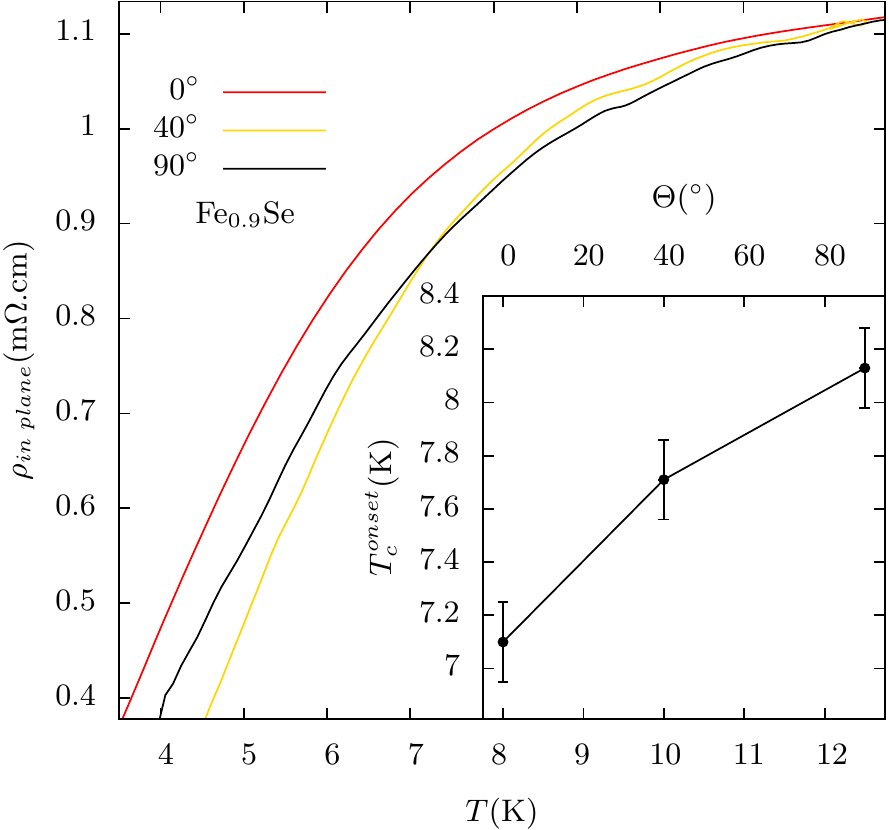}
\end{center}
\caption{Temperature dependence of the resistivity of Fe$_{0.9}$Se with $H$=14\,T for three different angles ($\Theta$=0\,$^\circ$, 40\,$^\circ$ and $90\,^\circ$). \emph{Inset:} Critical temperature as a function of the angle, $\Theta$.}
\label{c48-cruce}
\end{figure}

In principle, one could relate the anomalous behavior of the dissipation in Fe$_\text{1-y}$Se either to intrinsic properties of the material or  to the microstructure related vortex dynamics. However, as confirmed by our measurements, $H_{c2}$ and $T_c^{onset}$ are intrinsic characteristics of the material that vary monotonously with the angle between $0\,^\circ$ and $90\,^\circ$ and only depend on the anisotropy of the material\cite{blatter}. Also, it should be mentioned that the multiband nature of the electronic properties of $\beta$-FeSe does not affect the angular dependence of $H_{c2}(T,\Theta)$ and $T_c^{onset}(H,\Theta)$\cite{lour2}.
 Related to the above, Figure \ref{c48-cruce} shows the temperature dependence of the resistivity for Fe$_{0.9}$Se with $H$=14\,T at $\Theta=0\,^\circ$, $\Theta=40\,^\circ$ and $\Theta=90\,^\circ$. The critical temperature has a monotonic behavior with angle, see inset of figure \ref{c48-cruce}. Then as $T_c^{onset}$ and $H_{c2}$ have the same dependence in $\Theta$, we can conclude that the anomalous behavior of the resistivity is governed by vortex dynamics.
Furthermore, the curve at $\Theta=40\,^\circ$ crosses the curve at $\Theta=90\,^\circ$ and the difference increases with decreasing temperature. This is another indication that it is related to vortex movement. The presence of correlated defects has more influence in the resistivity as the temperature decreases due to the reduction of vortex mobility.

\begin{figure}[t]%h de here, b de bottom y t de top
\begin{center}
\includegraphics[width=8cm]{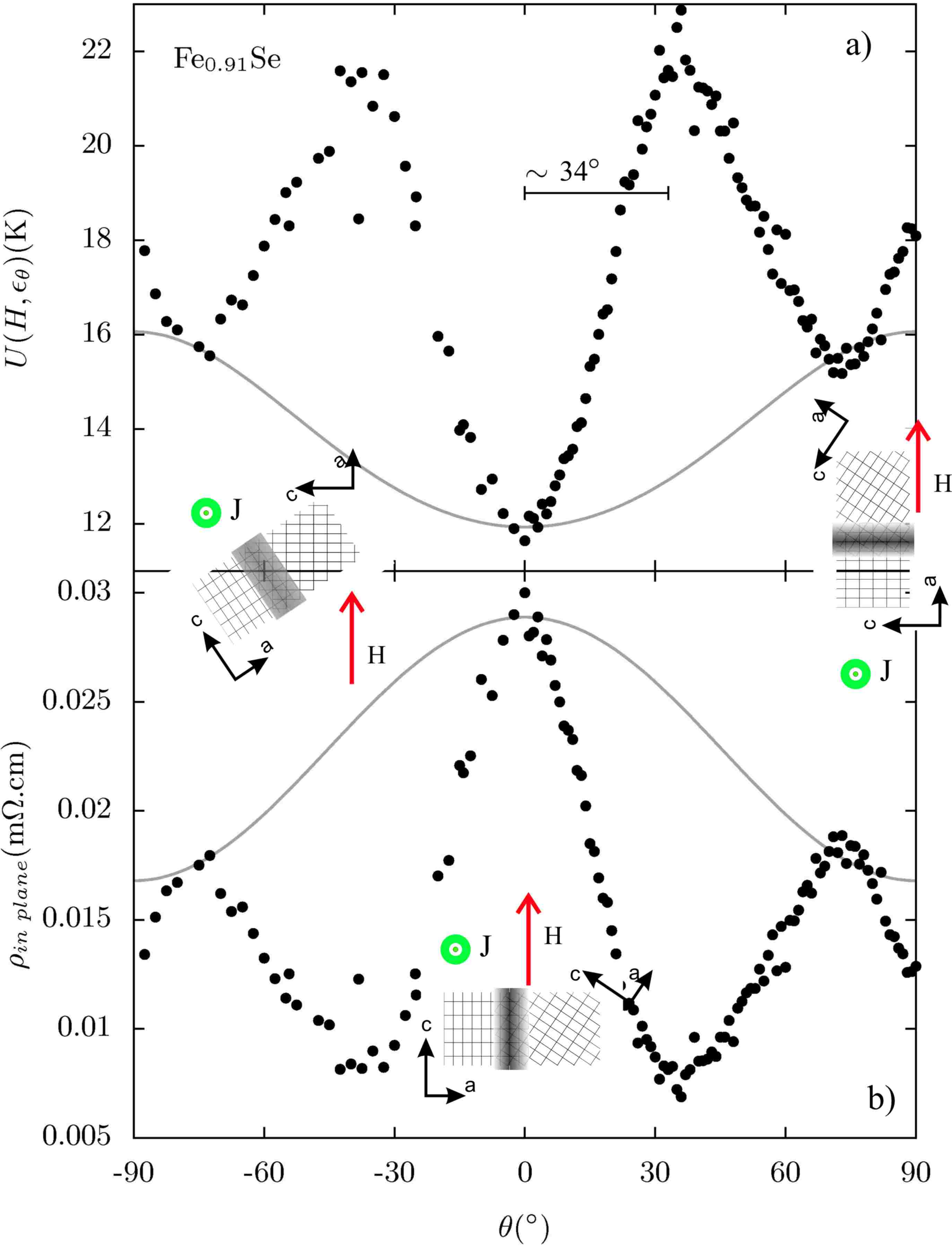}
\end{center}
\caption{a) Angular dependence of the activation energy and b) Angular dependence of the in plane resistivity, both for a single crystal of Fe$_{0.91}$Se for $H$=16\,T and $T$=4.90(3)\,K. The solid line represents the expected behavior for an anisotropic material, with $ \gamma$=1.20(5), in the presence of only point defects.
%% $Inset:$ $\theta$ is the angle between the normal to the plane and the applied magnetic field, $H$. The current, $I$, is always perpendicular to $H$.
The sketches represent the $a$  and $c$ axis direction with respect to the current and applied field.
}\label{dep-ang-c4}
\end{figure}

As presented in the previous section, the dissipation is well described by an Arrhenius law. Figure \ref{dep-ang-c4} presents the angular dependence of the activation energy for Fe$_{0.91}$Se. The data exhibit two contributions to the activation energy. One comes from point defects and is represented by the solid line. The other at $\Theta \sim \pm 34\,^\circ$ comes from a possible structure of correlated defects which act as pinning centers. To understand this contribution and the difference with Te doped samples, we focus on the crystal structure of our samples.

\subsubsection{Microstructural angular investigation.}\label{microstructure angular}
\begin{figure}[t]%h de here, b de bottom y t de top
\begin{center}
\includegraphics[width=0.5\textwidth]{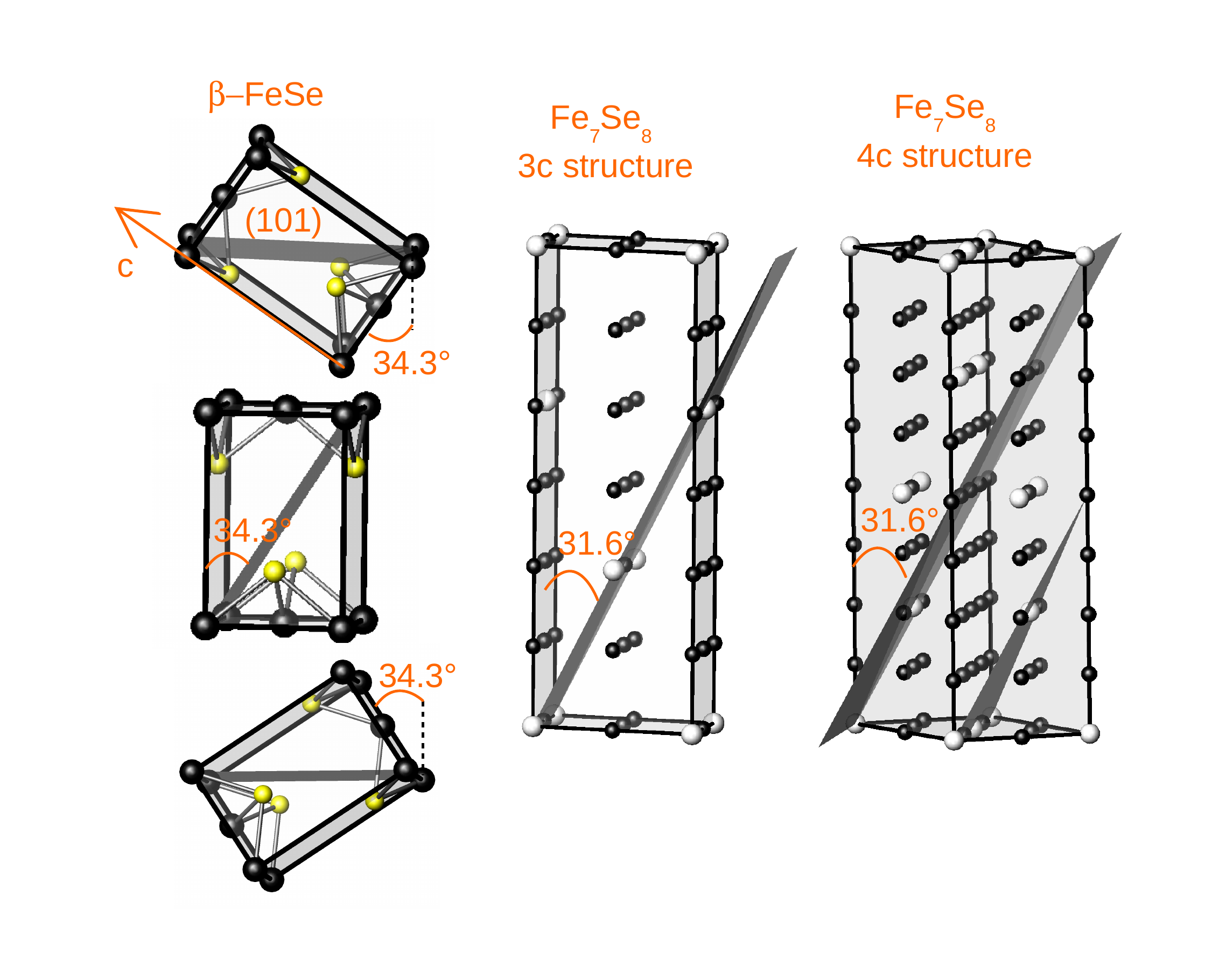}
\end{center}
\caption{Drawing of the crystal structure of our samples. Unit cell of the tetragonal phase ($Left$) with the different orientations of the $c$ axis and the unit cell of the hexagonal phase ($Right$) for the two possible superstructures 3$c$ and 4$c$. For the hexagonal phase we only show the Fe atoms in black and the vacancies in white.
}\label{estructura}
\end{figure}

\hfill\\
In the XR diffraction patterns we identified the  Fe$_\text{1-y}$Se $(00l)$ and $(l0l)$ families and the $(00l)$ one for the Fe$_7$Se$_8$ spurious phase, see figure \ref{Rx}.  
This mixture of phases and the interfaces between them could be the microscopic origin of the correlated defects acting as pinning center in Fe$_{0.91}$Se. As we already said, there is a nanoscale intergrowth of both phases and its shape and size depends on the crystal growth parameters\cite{wittlin}. In the hexagonal phase, the vacancies are ordered as a superstructure. Due to the crystal growth procedure, the most likely superstructure is the one called $3c$, in which the lattice parameter of the superstructure is 3 times that of the normal cell\cite{c3}. But also the $4c$ structure could be present, as we show from neutron diffraction data in the Appendix. In either case there is a dense plane of vacancies at an angle $\sim 32\,^\circ$ of the $c$ axis, i.e. the plane $(l0l)$. On the other hand, the tetragonal phase has twofold orientation of the $ab$ plane. One is parallel to the crystal surface and the other is at $\pm 34(1)\,^\circ$ to the normal of the surface, considering the measured lattice parameters $a=3.79 (1)\,\text{\AA}$ and $c=5.52(1)\,\text{\AA}$. Unit cells for the different phases of ours crystals are shown in figure \ref{estructura}. Therefore, the $ab$-plane of the tetragonal phase is close to the  most dense plane of vacancies of the $3c$ or $4c$ structure of the hexagonal Fe$_7$Se$_8$ phase. This could be acting as seed for both $\beta$ phase orientations during growth and it is a possible correlated defect. Additionally, when the magnetic field is applied in the direction  $\Theta = \pm 34(1)\,^\circ $, the vortices are crossing the $ab$-plane in their movement. Using the angular dependence of the dissipation described above, we know that the resistivity has a minimum when the vortices cross the $ab$-plane.  

\begin{figure}[h!!]%h de here, b de bottom y t de top
\begin{center}
\includegraphics[angle=0,origin=c, width=8cm]{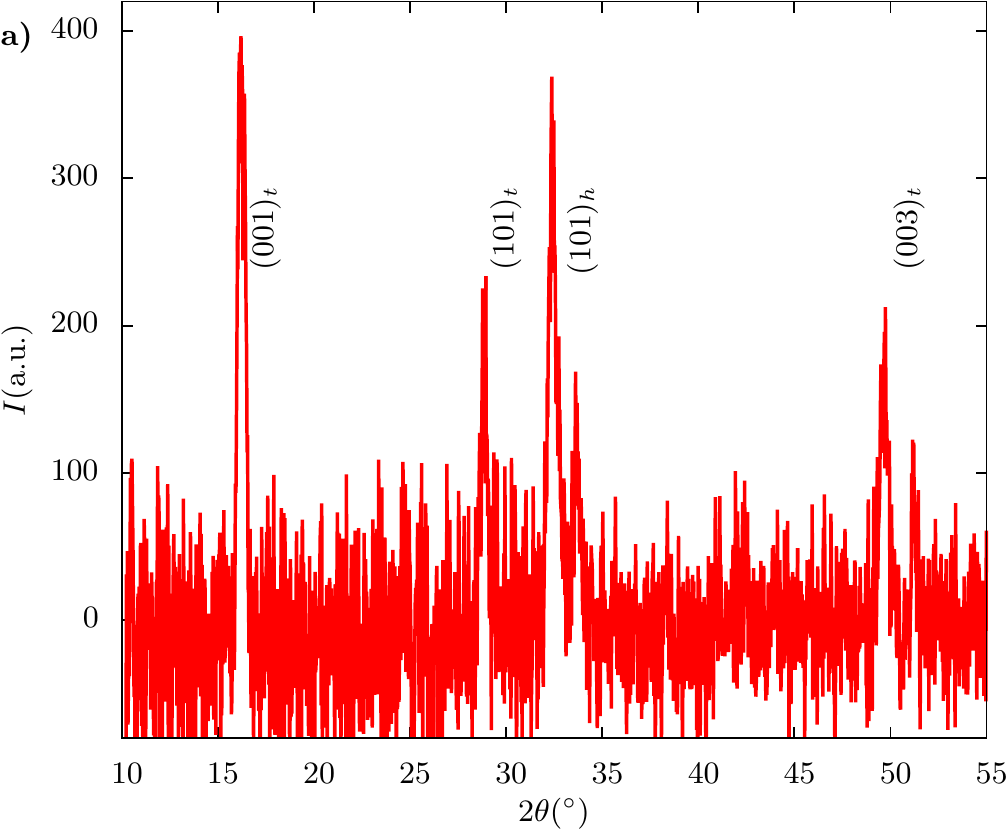}
\includegraphics[width=8cm]{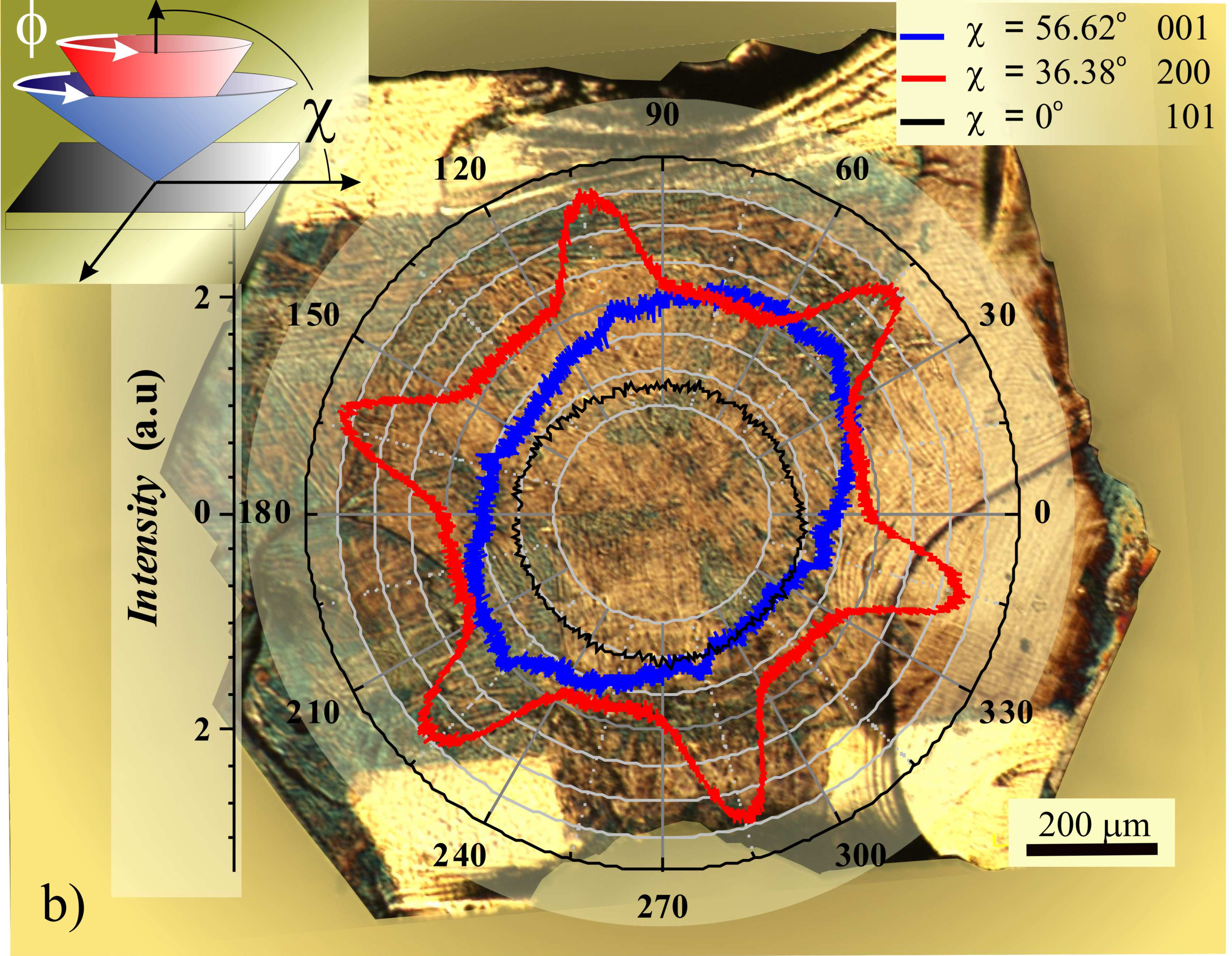}
\end{center}
\caption{a) XR-diffraction pattern taken at $\chi$=55.62\,$^\circ$  and $\phi$=43.7\,$^\circ$. The Miller indices of tetragonal phase and the hexagonal phase are indicated. b) Polar graphs of the XR diffraction $\phi$ scans patterns for the sample in a) taken at different values of the azimuthal angle $\chi$ coincident with the $(001)$, $(200)$ and $(101)$ reflexions of the $\beta$ phase indicated with blue, red and black lines superimposed to the sample photograph. The sketch in the upper inset indicates the geometry of  $\chi$  and  $\phi$ angles. }\label{ChiPhi}
\end{figure}
In figure \ref{ChiPhi}a) we show an XR diffraction pattern, fixing the diffraction condition at $\chi$=55.62\,$^\circ$, $\phi$=43.7\,$^\circ$. The sketch in the inset of figure \ref{ChiPhi}b) shows the $\chi$ and $\phi$ angles geometry. In this condition we see that the $ab$ plane and the (101) peak of the tetragonal phase are parallel to the (101) peak of the hexagonal phase. This is consistent with the dense plane of vacancies being the interface with the tetragonal phase in spite of the mismatch of 2\,$^\circ$ between the planes of both phases.

Microscopically, this boundary matching both phases should have just one direction and the minima in the dissipation as a function of the angle $\Theta$  should not be symmetric with respect to the $\Theta$=0\,$^\circ$ direction. We have found this situation only in very small samples with a  platelet surface  of 2500\,$\mu$m$^2$  or less. For larger samples, as that shown in figure \ref{ChiPhi}b), the pinning structure is symmetric and  we have found the resistance minima at $\Theta=\pm 34(1)\,^\circ$ for different directions of the current and contact arrangements as long as a constant Lorentz force is maintained. These large samples have a characteristic platelet  surface topography, namely that observed in figure \ref{ChiPhi}b) that shows lines with a sixfold symmetry.
To understand the boundary on the plane of the crystal and the two symmetrical minima on the angular dependence of the resistivity we include, in figure \ref {ChiPhi}b), $\phi$ angle scans at three $\chi$ angles. The pattern corresponding to $\chi$=0 and the $(101)$ diffraction shows the background. From the measured patterns in a $\phi$ scan we observed the $(001)$ and the $(200)$ diffractions, of the tetragonal $\beta$-phase, corresponding to the $c$-axis and $a$-axis directions respectively displaying a six fold geometry. Therefore, the intergrowth of the hexagonal phase is modifying the rotational symmetry of the tetragonal phase.

\begin{figure}[t]%h de here, b de bottom y t de top
\begin{center}
\includegraphics[width=8cm]{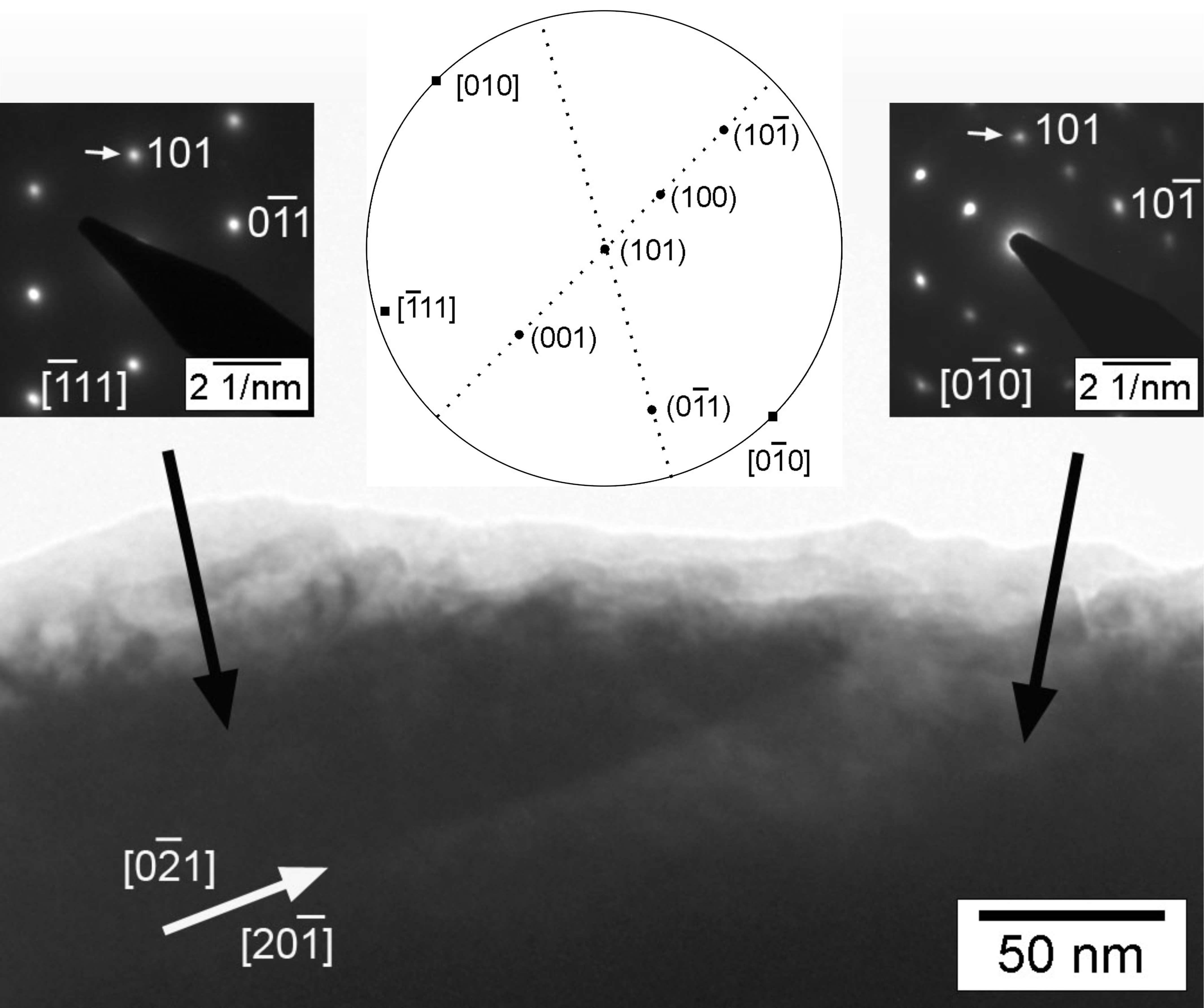}
\end{center}
\caption{ Transmission electron microscopy image of the a Fe$_{0.9}$Se crystal from  a  cross-section view. The diffraction patterns corresponding to the two observed orientations of the beta-phase have in common the 101 spot . The sketch indicates the  stereographic  projection  of the tetragonal structure centered on the normal of the (101) plane.
}\label{TEM}
\end{figure}

 For further assessment, transmission electron microscopy  images were taken from cross-section samples.
In Figure \ref{TEM} the bright field image including an interface between two orientations of the tetragonal $\beta$-phase can be observed. The corresponding selected area electron diffraction patterns show that the $\langle$111$\rangle-$type zone axis of the crystal on the left side is parallel to the $\langle$010$\rangle-$type zone axis of the crystal on the right side, having in common the 101 spot indicated by an arrow. The correspondence between the two crystals can be rationalized from the stereogram, sketched in the figure,  centered on the normal of the crystal surface, and oriented to match one of the six-fold geometry orientations of the tetragonal structure shown in figure \ref{ChiPhi}b). It can be observed that there are 120$\,^\circ$ between [0$\bar{1}$0] and [$\bar{1}$11] directions corresponding to the zone axis diffraction patterns observed in fig \ref{TEM}. Therefore, the two orientations observed by TEM are in accordance to the many-fold orientations observed by XR diffraction. The trace of the interface is close to the $\langle$201$\rangle$-type direction in both tetragonal structures and is indicated by a white thick arrow in Fig \ref{TEM}. 
Although the presence of the hexagonal phase was not observed at the interface of these two tetragonal domains, the angle between them is consistent with six-fold symmetry depicted in large samples as that of figure \ref{ChiPhi}b).

As a final remark, we want to go back to the very different scenario observed in the angular dependence of the resistivity on the samples with Te in figure \ref{dep-ang}. The absence of the extra resistivity minimum at $\Theta=\pm 34(1)\,^\circ$ indicates that there are no correlated defect in this case. As seen in figure \ref{MvsT}, the magnetic order, that is related with the vacancy order, is rapidly diluted with the inclusion of Te. Now there is no dense plane of vacancies in the hexagonal phase and therefore there is not a clear interface between the two phases. This has as a consequence that the angular dependences of the resistivity are similar of those of only point defects.

\section{Conclusion}\label{conclusion}

We have presented  the angular dependence of the dissipation in the superconducting state  of  FeSe and Fe(Se$_\text{1-x}$Te$_\text{x}$)   using  crystalline  materials with Fe$_\text{1-y}$(Se$_\text{1-x}$Te$_\text{x}$) intergrowth. This intergrown material is a solid solution from Fe$_{ 0.875}$Se to Fe$_{0.667}$Te.
We found that correlated defects are revealed as a hallmark of the angular dependence, in particular near the Se reach end.
 The in plane resistance in the superconducting state as well as the activation energy of the thermally activated response are consistent with the presence of correlated defects. These correlated defects act as pinning centers the vortices  in  very well defined angles, $\pm$34\,$^\circ$ from the crystal's normal, in the case of Fe$_\text{1-y}$Se. By adding Te, these defects disappear and there is not a clear indication of them in the angular dependence.
We have  presented an analysis of the microscopic nature of these defects based in the structure of the  intergrown materials, which are the superconducting  $\beta$-Fe(Se$_\text{1-x}$Te$_\text{x}$) and the iron deficient hexagonal  phase Fe$_\text{1-y}$(Se$_\text{1-x}$Te$_\text{x}$). The latter Fe deficient phase is magnetic near the Se reach end and was found to be a hexagonal material with two set of vacancies arrangements as related to the observed magnetic bulk properties as well as in the neutron diffraction experiments.
The existence of intergrowth of this phase favored the presence of two main orientations of   $\beta$-Fe(Se$_\text{1-x}$Te$_\text{x}$) in the crystals, and apparently plays a key role for the origin of  the correlated defects structure. The matching of both atomic structures define the  growth habit of the crystalline material as well as the correlated defects orientation.

Furthermore, information on intrinsic superconducting properties was provided. We reported an increase of the critical superconducting field with the Te content and a  maximum in the critical temperature for Te content x$\simeq$0.4 . An increase of the $T_c$ was observed with respect to crystals with only the $\beta$ phase, may be related with tensions on the interphase between the tetragonal and the hexagonal phases. For the resistivity in the normal state, we reported a maximum at a temperature that increases with the Se content.

\section{Acknowledgements}
We thank F. Gennari for help in sample preparation. 
Work partially suported by Conicet PIP 2014-0164, ANPCyT, PICT 2014-1265 and Sectyp U. N. Cuyo.  
Funding of this research by Spanish MINECO and FEDER programme (Project ENE2014-52105-R-9) is  gratefully acknowledged. Neutron diffraction experiments conducted at the LLB, were supported through the research proposals 11844, 12353 and 12354.

\appendix
\section*{Appendix}
\setcounter{section}{1}
\subsection*{Neutron diffraction analysis of Fe$_{1-y}$Se polycrystalline material}
Neutron powder diffraction was used to characterize the vacancy superstructure of the hexagonal phase Fe$_{7}$Se$_{8}$ and to investigate possible strain effects due to the mixture of phases.
 We performed the neutron powder diffraction experiment at the Laboratoire Le\'on Brillouin in the G4.1 instrument with $\lambda$=2.426\,$\text{\AA}$.  We also measured at the high resolution instrument 3T2 with $\lambda$=1.960\,$\text{\AA}$ to study the structural transition of the tetragonal phase.
 We compared two synthesized polycrystalline samples of FeSe, one with pure $\beta$-phase, and the other with a mix of tetragonal and hexagonal phases in a similar percentage as the crystals studied in this  work. 

\begin{figure}[b]%h de here, b de bottom y t de top
\begin{center}
\includegraphics[width=8cm]{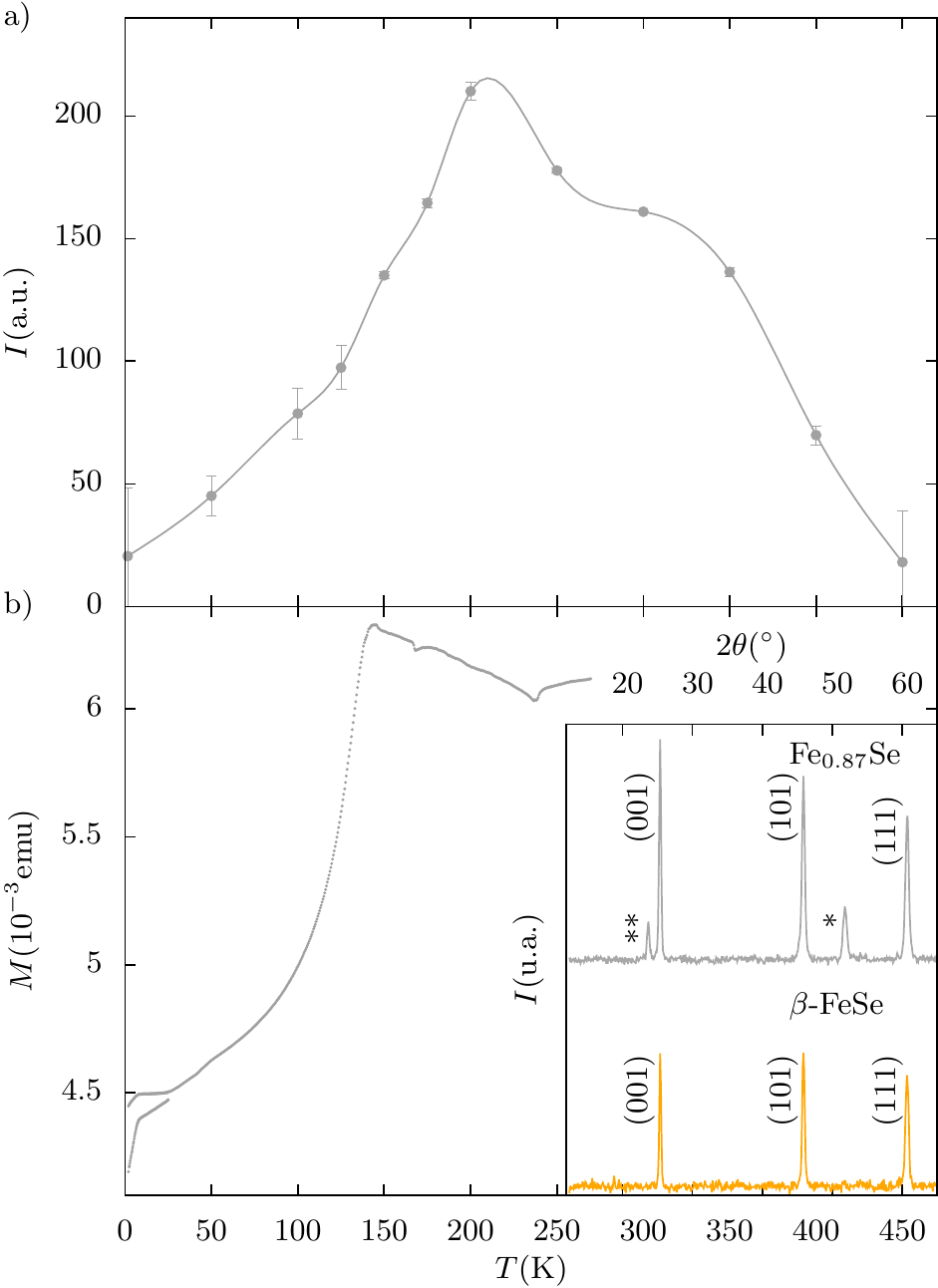}
\end{center}
\caption{a) Temperature dependence of the integrated intensity of the magnetic reflection (001) in the 2$\theta$ range  21.7$^\circ$ - 24.3\,$^\circ$ for Fe$_\text{1-y}$Se, the line is a guide to the eye. b) Magnetization for Fe$_\text{1-y}$Se with an applied magnetic field of 20\,Oe. $Inset:$ Neutron diffraction patterns for Fe$_\text{1-y}$Se and $\beta$-FeSe for $T=250$\,K. The tetragonal phase Miller indices are indicated in brackets. $\ast$ indicates peaks corresponding to the hexagonal phase and $\ast \ast$ indicates the magnetic peak (001)  of the hexagonal structure.}\label{IvsTh}
\end{figure}
The polycrystals were obtained as described in Section \ref{crystal growth} using a maximum temperature of   680\,$^\circ$C  for 48 hours and of 437\,$^\circ$C for 1 month for the samples with mixture of phases and with pure $\beta$-phase respectively. 
 The analysis of the neutron diffraction pattern revealed the mixture of tetragonal and hexagonal phases in the first  case that we will refer as Fe$_\text{1-y}$Se and the presence of only the tetragonal phase $\beta$-FeSe in the other, see inset of  figure  \ref{IvsTh}. 

Figure \ref{IvsTh} a) shows the integrated intensity of the magnetic reflection (001), marked with $\ast \ast$ in the inset, as a function of the temperature. This gives information about  the vacancy order superstructure, present in 
Fe$_\text{1-y}$Se. 
In the pure $4c$ superstructure, a gradual drop below 210\,K has been reported\cite{japoneses} for the (001) magnetic reflection. Nevertheless, for the $3c$ superstructure, the same magnetic reflection shows a sudden drop, indicative of a spin reorientation around 140\,K (recall that magnetic peaks relate to the projection of the magnetic moment on the plane perpendicular to the given direction). In agreement with the above, the behavior observed in Fig.\ref{IvsTh}a) indicates a mixture of the $3c$ and $4c$ phases.
The behavior of the bulk magnetization, shown in Fig.\ref{IvsTh}b), displays a steep increment with increasing temperature, with a maximum at 140 K. This trend may be reconciled with the neutron diffraction analysis also in terms of the coexistent $3c$ and $4c$ superstructures. The magnetic subnetwork related to the $4c$ structure smoothly starts reorientation towards the (001) direction when the temperature  decreases below 210\,K, getting closer to the easy axis of the $3c$ structure. The neutron diffraction intensity diminishes, but the bulk magnetization increases. On the other side, the $3c$ subnetwork undergoes a sudden reorientation at around 140\,K, and at this point the bulk magnetization quickly falls with decreasing temperature.
This is consistent with a mixture of $3c$ and $4c$ superstructures\cite{terzieff1978antiferromagnetic}. The magnetization of the crystalline samples, figure \ref{MvsT}a) in Section \ref{crystal growth} shows that it probably has the same mixture of superstructures.

\begin{figure}[t]%h de here, b de bottom y t de top
\begin{center}
\includegraphics[width=8.5cm]{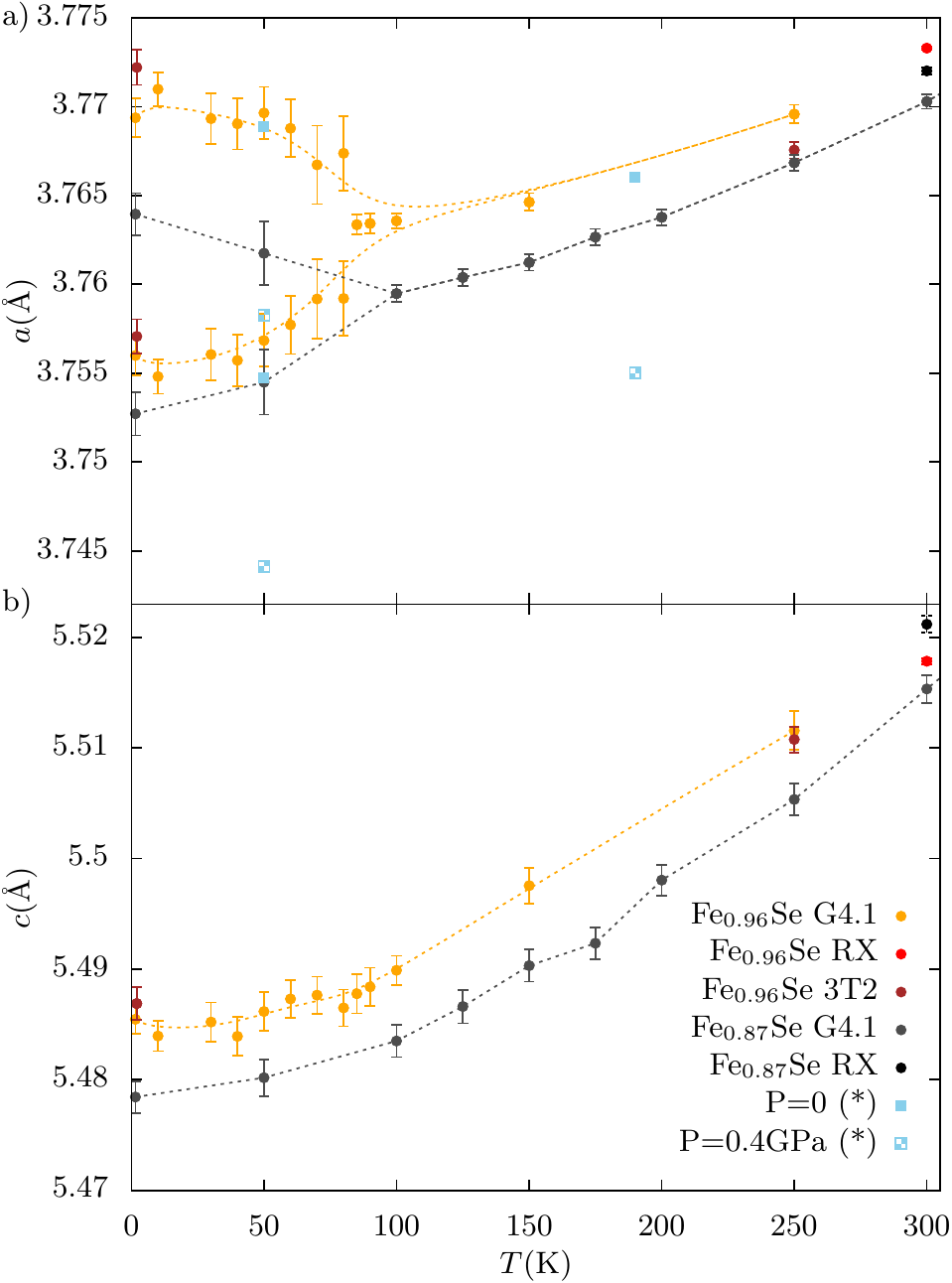}
\end{center}
\caption{a) Temperature dependence of the lattice parameter $a$ in the tetragonal phase and $a/\sqrt{2}$ and $b/\sqrt{2}$ in the orthorhombic phase and b) of the $c$ axis parameter for $\beta$-FeSe and Fe$_\text{1-y}$Se. The symbols correspond to XR or neutron diffraction data as indicated, also data marked with $\ast$, taken from reference \cite{millican2009pressure} with and without applied pressure are plotted for comparison.}\label{a-c-neutrones}
\end{figure}

From the analysis of the neutron diffraction data, we also obtained the temperature dependence of the lattice parameters of the $\beta$-phase for the two polycrystals under study, see figure \ref{a-c-neutrones}. In both cases, there is a structural transition at $\sim$90\,K from tetragonal $P4/nmm$ to orthorhombic $Cmma$. There is a good agreement between the lattice parameters obtained with neutron diffraction at the G4.1 instrument, the 3T2 instrument and with XR measurements. On the other hand, there is a small difference between the lattice parameters obtained from the $\beta$-FeSe and the sample with mixture of phases. 

\begin{figure}[t]%h de here, b de bottom y t de top
\begin{center}
\includegraphics[width=8cm]{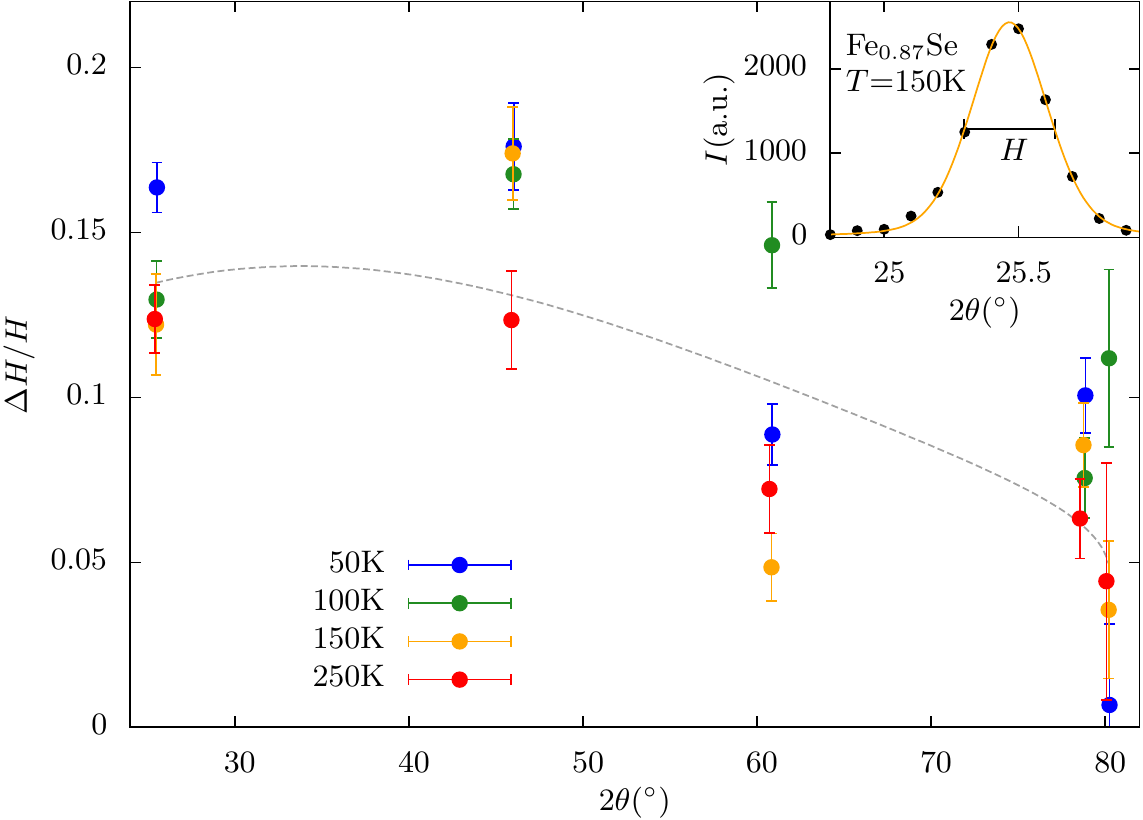}
\end{center}
\caption{Normalized difference peak width, $\frac{\Delta\text{H}}{\text{H}}=\frac{2\text{(H(Fe$_\text{1-y}$Se)-H($\beta$-FeSe))}}{\text{H(Fe$_\text{1-y}$Se)+H($\beta$-FeSe)}}$ for the different reflections and temperatures. $Inset:$ Example of the fit of the Pseudo-Voight function for the (001) reflection for Fe$_\text{1-y}$Se at $T$=150\,K.}\label{DH-H}
\end{figure}

This difference is not enough to explain the increment of transition temperature in the resistivity described in  Section \ref{superconducting}.  Given the sensitivity to external pressure, it was reported that a change in  $ T_{c}$ of $\sim$3.5\,K is obtained with a pressure of 0.4\,GPa\cite{bendele2012coexistence}. The difference of the lattice parameter $a$ obtained in reference \cite{millican2009pressure} at 0 and 0.4\,GPa is larger than the difference obtained for the two samples under study, as shown in figure \ref{a-c-neutrones}. 
A systematic microstructural observation accompanies the variation in the critical temperature in our resistivity data. In this case, broader peaks in the diffraction pattern of the sample with mixture of phases are expected. To confirm this, we fitted a Pseudo-Voight function to each peak of the tetragonal phase with the lorentzian and gaussian weight defined by the Rietveld refinement of the whole diffractogram at that temperature. Figure \ref{DH-H} shows the ratio between the difference and the average of the measured tetragonal peak width in Fe$_\text{1-y}$Se and $\beta$-FeSe. For all temperatures and reflections, the polycrystal with mixture of phases presented  wider peaks. This 
could be interpreted in terms of tensions in the sample with mixture of phases (probably at the boundary regions), defective structures, grain size disorder, etc.

\bibliography{iopart-num1}{}
\bibliographystyle{iopart-num}

\end{document}